\documentclass[superscriptaddress,showpacs,aps,twocolumn,prb,floatfix]{revtex4}
\usepackage[utf8]{inputenc}
\usepackage{amssymb}
\usepackage{amsmath}
\usepackage[dvips]{graphicx}
\usepackage{subfigure}
\usepackage{bm}
\usepackage{color}
\usepackage{soul}
\usepackage{gensymb}
\usepackage{wasysym}
\usepackage{xcolor}
\usepackage[export]{adjustbox}

\begin{document}
 
\title{2D superconductidy driven by interfacial electron-phonon coupling at the BaPbO$_3$/BaBiO$_3$ bilayer}
\author{S. Di Napoli}
\affiliation{Departamento de F\'{\i}sica de la Materia Condensada, GIyA-CNEA, Av. General Paz 1499, (1650) San Mart\'{\i}n, Pcia. de Buenos Aires, Argentina}
\affiliation{Instituto de Nanociencia y Nanotecnolog\'{\i}a (INN CNEA-CONICET), 1650 San Mart\'{\i}n, Argentina}

\author{C. Helman}
\affiliation{Instituto de Nanociencia y Nanotecnolog\'{\i}a (INN CNEA-CONICET), 1650 San Mart\'{\i}n, Argentina}
\affiliation{Centro Atómico Bariloche and Instituto Balseiro, CNEA, 8400 S. C. de Bariloche, Argentina}

\author{A.M. Llois}
\affiliation{Departamento de F\'{\i}sica de la Materia Condensada, GIyA-CNEA, Av. General Paz 1499, (1650) San Mart\'{\i}n, Pcia. de Buenos Aires, Argentina}
\affiliation{Instituto de Nanociencia y Nanotecnolog\'{\i}a (INN CNEA-CONICET)}

\author{V. Vildosola}
\affiliation{Departamento de F\'{\i}sica de la Materia Condensada, GIyA-CNEA, Av. General Paz 1499, (1650) San Mart\'{\i}n, Pcia. de Buenos Aires, Argentina}
\affiliation{Instituto de Nanociencia y Nanotecnolog\'{\i}a (INN CNEA-CONICET), 1650 San Mart\'{\i}n, Argentina}

\email{vildosol@tandar.cnea.gov.ar}

\begin{abstract}
The recent discovery of 2D superconductivity  at the interface of BaPbO$_3$ (BPO) and BaBiO$_3$ (BBO) has motivated us to study in depth the electronic and structural properties and the relation between them in this particular heterostructure, by means of first-principles calculations.  Our results indicate that the breathing distortions, the charge ordering and the semiconducting behaviour that characterize the parent compound BBO in its bulk form, are preserved at the innermost layers of the BBO side of the BPO/BBO bilayer. On the other hand, at the interface, there is a partial breaking of the breathing distortions with a concomitant charge transfer between the interfacial Bi ions and the on top BPO layer.  We
show that two types of carriers coexist at the interface, the delocalized 3D like sp states coming from Pb ions and the quasi 2D s states from the Bi ones. We obtain a substantial electron-phonon coupling between the 2D Bi states with the interfacial stretching phonon mode and a large density of states that can explain the critical temperature experimentally observed bellow 3.5 K. We hope these findings will motivate future research to explore different interfaces with charge ordered semiconductors as BBO in order to trigger this fascinating 2D behavior.

\end{abstract}

\maketitle

In the last decades, the scientific community has devoted a great effort to find new emerging properties at the interfaces of oxide heterostructures with the aim of diversifying the functionalities of existing electronic devices. 

In oxide materials, the covalency of the bonds and the fact that the crystal structures  are conducive to vary the chemical composition,  enable the possibility of controlling the different electronic degrees of freedom, the interrelation among them and/or with the crystal structure and its dynamics. This is the clue to design new devices with novel technologies in which the interface between the constituent materials plays a fundamental role.   
Different physical phenomena such as long range orders, superconductivity or metal-insulator transitions can be exploited for this purpose.  

In particular,  the discovery of interfacial superconductivity in oxide heterostructures has caught much attention. First, it was demonstrated that superconductivity can occur at the interface between two band insulators such as  LaAlO$_3$ and SrTiO$_3$  \cite{Reyren1196}. Later, it was the turn of the strongly correlated cuprates La$_2$CuO$_4$ and  LaSrCuO$_4$, for which high-T$_c$ interfacial superconductivity was reported \cite{Gozar2008} and that it can occur even at a sole CuO$_2$ plane \cite{Logvenov699}. And very recently, 2D superconductivity was probed at the FeSe/SrTiO$_3$ interface \cite{PhysRevLett.124.227002} independently of the thickness of the FeSe film.

Interestingly, 2D superconductivity was also observed at the BaPbO$_3$/BaBiO$_3$ bilayer~\cite{Meir2017}. Separately, BaBiO$_3$ (BBO) is a charge ordered Peierls-like semiconductor while  BaPbO$_3$ (BPO) films are metallic, but the emerging 2D superconductivity remains unclear. The 2D superconducting T$_c$ reported for the BPO/BBO bilayer reached up to  ~3.5 K depending of the thickness of the BPO layer. In this work, we give insight into the detailed electronic structure, structural reconstructions and interfacial vibrations of this bilayer by means of first-principles calculations. 

\begin{figure*}[t]
\includegraphics[width=2.\columnwidth]{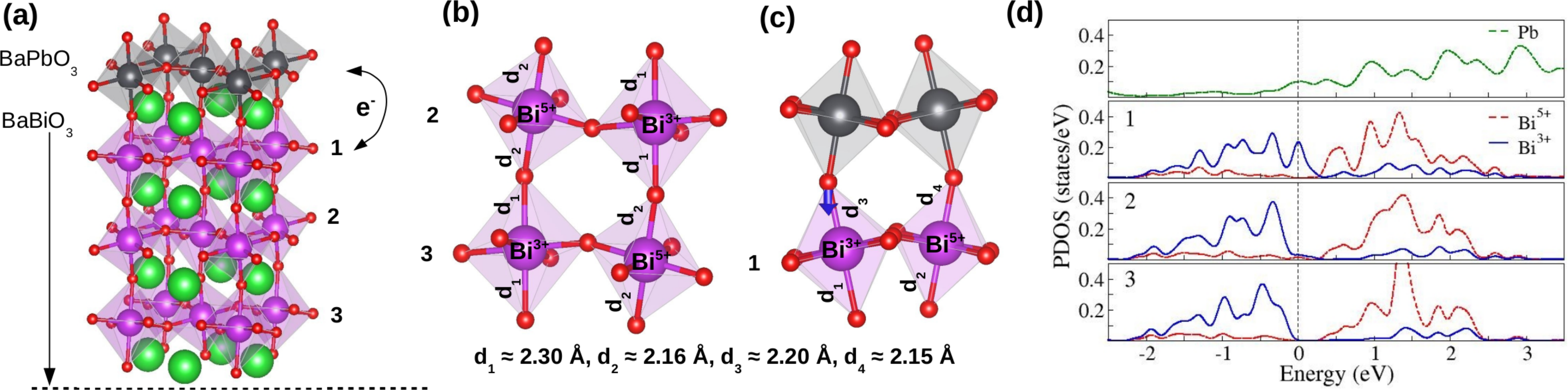}
\caption{(a)Schematic representation of the BaBiO$_3$/BaPbO$_3$ interface. We show half of the structure used in our simulations. Numbers 1,2,3 represent the plane number. (b) and (c) Octahedral environment of Bi atoms belonging to the bulk and interface regions, respectively. (Green:Ba, magenta:Bi, red:O and grey:Pb atoms).(d) Bi (s) and Pb (sp) partial density of states (PDOS) projected onto the planes labeled in (a).}  
\label{fig:structure}
\end{figure*}

The phase diagram of bulk BBO is very rich and deserves a brief description in order to understand the properties at the interface. Among the most outstanding features is the fact that it becomes superconducting under hole doping in the vecinity of a charge-density wave ordered phase. Despite the absence of magnetic instabilities in the parent compound, the critical temperature (T$_c$) can be as high as ~32 K \cite{Nature-babio3} for Ba$_{0.6}$K$_{0.4}$BiO$_3$ (BKBO) when K replaces Ba sites which are not in the BiO$_2$ superconducting plane, or it can reach up to ~13 K  \cite{Tc-bapbbio3} for BaPb$_{0.7}$Bi$_{0.3}$O$_3$ (BPBO) when Pb substitutes Bi. 

Undoped BBO crystallizes in a double perovskite type structure. The formula unit forces the Bi atom to have the formal valence +4, however due to the lone pair effect, this valence state is unstable so that, at low temperatures, the system develops an insulating charge ordered state with alternated Bi$^{+5}$-Bi$^{+3}$ sites. In reality, the difference in charge between both Bi sites is much less than the formal valence due to hybridization and screening effects\cite{PhysRevB.28.4227,PhysRevB.81.085213}. Actually, Bi-O bonds can be described by spatially extended hybridized 6s-2p orbitals \cite{PhysRevB.26.2686,PhysRevLett.60.2681}. The charge disproportionation induces structural distortions characterized by the tilting of the octahedra and, specially,  by the breathing modes of the Bi-O bonds. As the temperature goes down, the system undergoes several transitions in the crystal structure loosing its symmetry, going through a cubic (T$>$750 K),  rhombohedral (405 K $<$ T $<$ 750 K) and monoclinic (T $<$ 405 K) phases \cite{Cox1976969}. Under hole doping (above a certain optimal value) there is no need to avoid the +4 valence state, the system turns simple cubic and metallic, and more important, superconducting bellow T$_c$. The physical nature of the superconducting phase has been debated during several decades. Recently, it has been demonstrated through angle-resolved photoemission spectra (ARPES) that the large electron-phonon coupling, in particular with those phonon modes related to the breathing distortions, can account for the high T$_c$ \cite{PhysRevLett.121.117002}. It has also been confirmed that it is necessary to go beyond local or semilocal functionals as LDA or GGA to account for the long range exchange interactions and that, density functional theory (DFT) calculations that take into account the non-local Hartre-Fock exchange interactions by means of screened hybrid functionals, can describe the physics correctly~\cite{Kotliar2013,PhysRevB.81.085213}.    

On the other hand, bulk BPO, the other compound of the bilayer, is monoclinic at low T  and pure BPO films were reported to be metallic \cite{2007-BaPbO3,Meir2017}. The valence $+4$ that imposes the perovskite formula unit for the Pb site implies that the 6s and 6p bands are practically empty. Different from what happens in BBO, there is only one crystal site type for Pb. It was suggested that the 2D superconductivity at the BBO/BPO interface was induced by the interfacial strain and that BPO acts as a dopant. 

In this work, we provide a microscopic description of the electronic structure at the interface of this bilayer and study the coupling of the interfacial electronic states with the lattice dynamics. We show that at the BPO/BBO interface there is a partial breaking of the BBO charge order which gives rise to 2D electronic states that present a strong coupling with interfacial phonons. This mechanism of generating 2D electronic states is similar to the one predicted for the Bi terminated (001) surface of pure BBO~\cite{Vildosola2013}. 

We perfom DFT calculations including a fraction of non-local Hartre-Fock exchange interactions through the  Heyd-Scuseria-Ernzerhof (HSE06) functionals~\cite{2003-HSE1,2006-HSE2} as implemented  in the Vienna \textit{ab initio} package (VASP)~\cite{VASP,PAW-VASP}. Technical details are given in the Supplementary Information (SI) file.

XRD studies reported in Ref. \citenum{Meir2017} indicate that BBO grows completely relaxed on SrTiO$_3$ substrates while there is a clear strained growth of the BPO layer on BBO. Based on this experimental evidence, we simulate the BPO/BBO bilayer considering a supercell in the  monoclinic crystal structure of BBO using the experimental lattice parameters\cite{Cox1976969} and relaxing the internal positions. The description of all the studied slabs BPOi/BBOj with different values of $i$ and $j$ is presented in the SI file. In this notation, the supercell is built along the (001) direction, $i$ is the number of BPO layers and $j$ the corresponding number for BBO. The supercell is rotated 45$^{\circ}$ with respect to the simple cubic single perovskite and contains two Bi or two Pb sites per layer. 

In order to properly describe the breathing distortions in BBO, a rather dense k-mesh is necessary\cite{PhysRevB.73.212106}. The relaxation of internal positions of the different slabs is performed within the Perdew-Burke-Ernzerhof (PBE)~\cite{PBE96} functional. In the SI file we show that PBE gives reasonable results for the breathing distortions and the charge disproportionation as compared to the HSE outcome. Its main drawback is the description of the bandwidths and the electronic bandgap so that we use the HSE functionals just at the electronic level for these large supercells. 

In Fig. \ref{fig:structure} a), we show half of the supercell considered for the BPO2/BBO6 slab. Layers indicated as 1, 2 3 correspond to half of the BBO part: layer 1 being the interfacial Bi layer while layer 3 is the innermost one.  In Figs. \ref{fig:structure} b), c) and d) the crystal and electronic reconstructions after relaxation of the internal parameters can be observed. Fig. \ref{fig:structure} b) shows a zoom of the Bi-O octahedra at layer 3. The  Bi$^{+3}$-O  and Bi$^{+5}$-O distances, d$_1$= 2.30~\AA\,\ and d$_2$=2.16~\AA, respectively, present very similar values to the ones obtained for bulk BBO (see Table S1 in the SI file). On the other hand, Fig. \ref{fig:structure} c) shows the crystal structure at the interface. The interfacial distance Bi$^{+3}$-O, d$_3$, shrinks about 5 \% with the respect to the innermost values d$_1$, as indicated with the blue arrow in the figure,  while the interfacial Bi$^{+5}$-O, d$_4$, remains practically unchanged. This partial suppression of the breathing distortion at the interface is in line with the electronic reconstruction that can be observed in the layer by layer projected density of states (PDOS) of Fig. \ref{fig:structure} d). At the top panel, the metallic sp-Pb PDOS is shown. At the botton, the s-Bi$^{+3}$ and s-Bi$^{+5}$ PDOS of layer 3 indicate that the slab holds the  semiconducting behavior away from the interface. At this layer, the s-Bi$^{+3}$ band is full while the s-Bi$^{+5}$ one is empty. Similarly, this charge disproportionation among the Bi sites survives at layer 2. However, at the interface, layer 1 turns out to be metallic due to a charge transfer between Bi$^{+3}$ and the on top BPO layer. Curiously, the s-Bi$^{+5}$ band persists empty. 

\begin{figure}[h]
\includegraphics[width=1.\columnwidth]{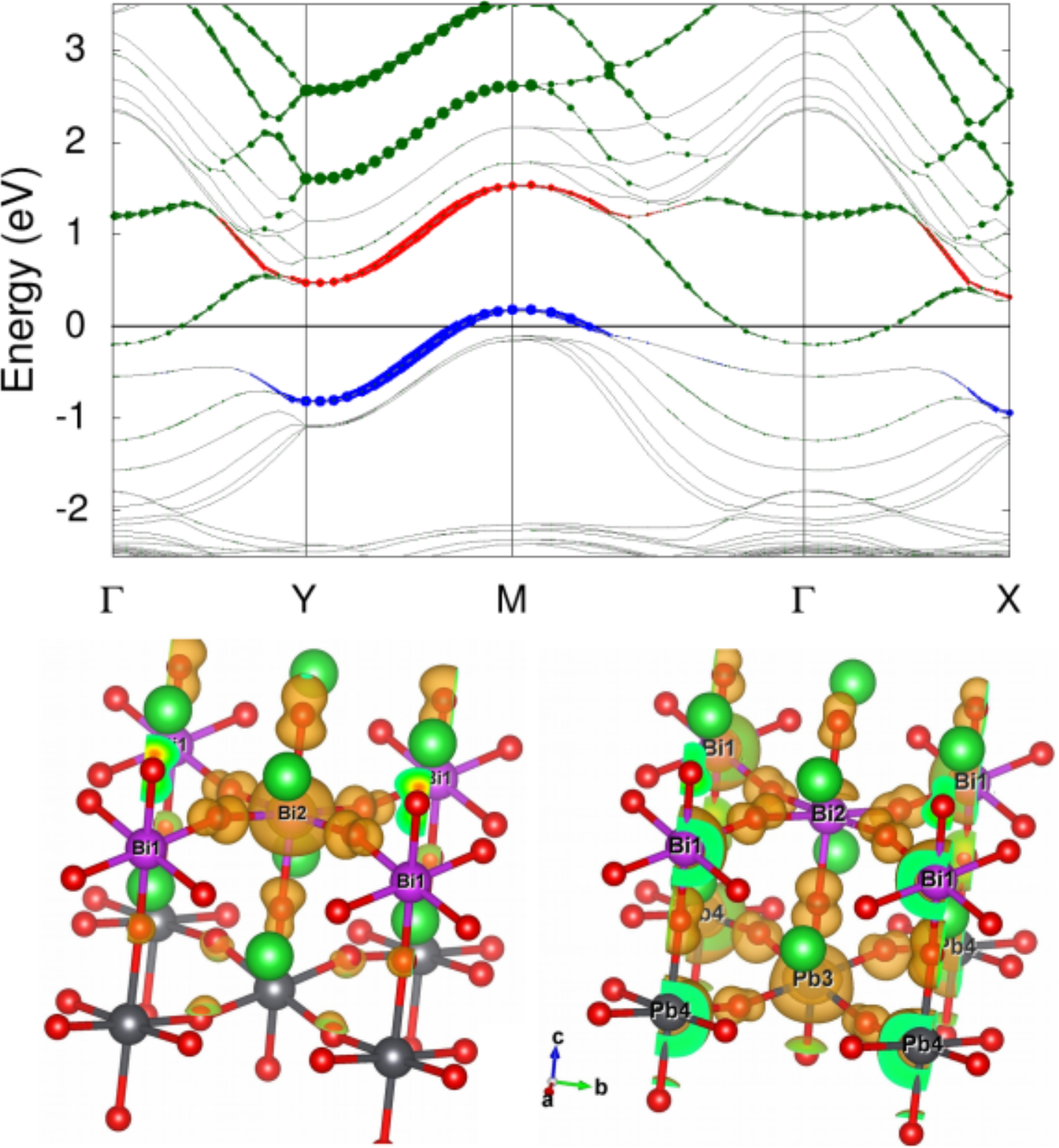} 
\caption{(top) HSE bandstructure of the BPO2/BBO6 heterostructure. The $s$-character of interfacial Bi$^{5+}$ (red circles), Bi$^{3+}$ (blue circles) and the $sp$ character of Pb (dark-green circles) are shown. (bottom) Charge density obtained from the states crossing the Fermi level from below (left) and from above (right) corrresponding to charge density mainly centered at the interfacial Bi$^{3+}$ and Pb (and Bi$^{5+}$) atoms, respectively. }
\label{fig:HSE-bands}
\end{figure}

Fig. \ref{fig:HSE-bands} (top panel) depicts the bandstructure calculated for the BPO2/BBO6 slab. Here, the s-Bi$^{+3}$ character of the Bi atom at layer 1 is highlighted in blue. The s-Bi$^{+5}$ character for the same layer is in red and the total sp-Pb character in green. It is important to remark that while the s-$Bi^{+3}$ states at Fermi energy (E$_F$) are quasi 2D, since only the states lying at layer 1 contribute to E$_F$, the sp-Pb bands are more delocalized and 3D like. The calculations done for the BPO4/BBO6 slab indicate that all the Pb atoms contribute equally to E$_F$. At the botton of Fig. \ref{fig:HSE-bands}, we show the calculated charge density for the bands crossing E$_F$. The charge density of the hole band with more s-Bi$^{+3}$ character is centered around the interfacial Bi$^{+3}$ sites while the electron band with sp-Pb character is clearly more delocalized with substantial weight at the oxygens of the BBO part.   

Interestingly, comparing the bandstructure of the BPO2/BBO4, BPO2/BBO6 and BPO2/BBO8 (see the SI file), the height of the hole pocket does not significantly change with the number of BBO layers, as long as they are more than 4. The reason is that the partial suppression of the charge order is a 2D effect at the interface. On the other hand, the doping effect of the BPO slab is a bit more sensitive with its width; the wider the BPO part the more effective is the doping on the BBO one. The size of both the hole and electron pockets increases with the width of the BPO side, as can be observed by comparing the BPO2/BBO6 with the BPO4/BBO6 in the SI material. We will come back to this point when we discuss the electron-phonon coupling.   

Both materials, BPO and BBO, have been predicted to have large topological gaps both in the electron and hole doping regimes \cite{Yan2013,Li2015}. In view of these properties, we have assessed the role of spin-orbit (SO) coupling on these interfacial effects of the bilayer. As shown in the SI material, we observe the opening of gaps when the SO is turned on, but these gaps are far away from the Fermi level of the interface studied here. Thus, we conclude that the topological properties of these materials do not affect the electronic behaviour of this bilayer.

We now study the coupling of these interfacial states with the lattice dynamics. Yin et. al. have shown that for doped BBO in its bulk form, the three phonon modes that couple stronger to the electrons are the breathing, the stretching (ST) and the ferroelectric (FE) modes\cite{Kotliar2013}.  At the BPO/BBO interface, the 3D breathing mode is obviously no longer operative and, therefore, we study the effect of the other two modes and estimate their contribution to the electron-phonon coupling.

As proposed in Refs. \citenum{PhysRevLett.55.837} and \citenum{PhysRevB.44.5388} and described in the SI file, the electron-phonon matrix elements can be estimated at a given wave vector q and phonon mode $\nu$, through a frozen phonon approach directly from the shift of the bands calculated for a proper supercell, as long as q is commensurate with the lattice.

In the optimally doped bulk BBO, the undistorted crystal is a single  simple cubic perovskite. As shown before, in the bilayer studied here, the breaking of charge order at the interface is partial so that the system remains in a double perovskite configuration. The simulated supercells allow us to estimate the electron-phonon matrix elements for the interfacial ST and FE modes and compare them with the bulk BBO ones, at the wave vector $\pi/a(1,1)$ of the single perovskite structure, to assess their relative strength. 

\begin{figure}[h]
\includegraphics[width=1.\columnwidth]{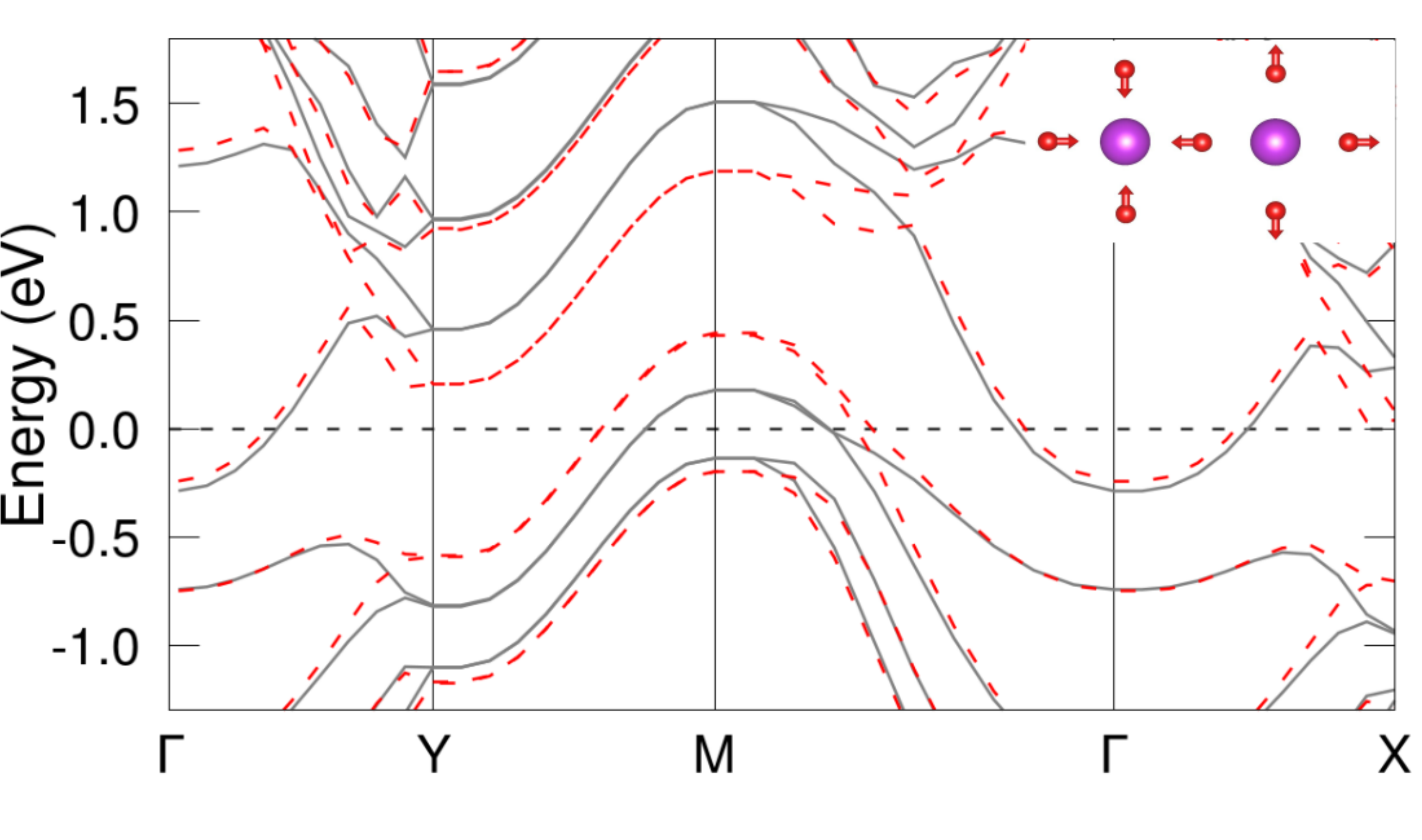}
\caption{HSE bandstructures of the BPO2/BBO4 heterostrusture in the undistorted (black solid line) and ST (dashed red lines) distortions. In the right insets, the  ST frozen phonon displacements are shown. The oxygen atoms's displacements considered are 0.044~\AA }.   
\label{fig:fr-ph-bands}
\end{figure}

The frozen phonon calculations are perfomed both with the PBE and HSE functionals (see the SI material for a detailed comparison). In Fig. \ref{fig:fr-ph-bands}, we plot the calculated HSE bandstructure of the BPO2/BBO4 slab in its relaxed structure (black solid line) together with the ones obtained for the ST (red dashed lines) distortions.  The effect of the FE interfacial distortion is negligible (see SI file). On the other hand, there is a substantial change in the electronic bands that cross E$_F$ for the ST mode, especially for the hole pocket surrounding the $M$ point that is the one with more s-Bi$^{+3}$ character. For the frozen-phonon distortions we consider a small atomic displacement of $|u|\sim 0.044\AA$ as in Ref. \citenum{Kotliar2013}. The other bands at $E_F$ come from the nearly empty Pb ones for which we haven't observed any sizeable electron-phonon coupling.

The estimated matrix element for the ST mode at the M point is 4.6 eV/$\AA$ with PBE, and turns to 5.9 eV/$\AA$ with HSE. The corresponding values of these shifts for the optimally doped bulk BBO are 5.1 and 8.9 eV/$\AA$ with PBE and HSE, respectively \cite{Kotliar2013}. Two things are important to remark. On one hand, the difference between the PBE and HSE for the interfacial ST mode is not as important as for the doped bulk BBO. This is probably due to the fact that at the interface, the breaking of charge ordering is only partial. Second, the interfacial ST electron-phonon matrix elements for the bilayer are substantial. They are smaller but of the same order as the strongest coupling in bulk BBO. It is noteworthy to say that we also obtain a value of ~4.6 eV/$\AA$ within PBE for the matrix element of the thicker BPO4/BBO6 slab. The main difference between the two slabs lie mainly the density of states (DOS) at $E_F$ per spin, $N^{\uparrow}_0$, that is larger for the thicker one, being 1.25 and 1.90 states/eV within PBE, respectively.

Then, the contribution to the electron-phonon coupling $\lambda$ from this interfacial stretching mode (IST), $\lambda_{IST}$, is obtained by performing the double delta sum in eq. S1 of the SI file, with the matrix elements calculated along the full Brillouin zone. Gaussian functions have been used to simulate the delta's. The phonon frequency of the IST mode, $\omega_{q IST}$ with $q=0$, has been estimated by a direct approach within the frozen phonon approximation, through which the changes in total energy are calculated in real space by slightly displacing the atoms from their equilibrium positions in the interfacial ST mode, giving $\omega_{0,IST} \sim 0.059$\,eV\,\footnote{We also confirm this value from the second derivatives of the energy with respect to atomic displacements (Hessian matrix) to obtain all the vibrational frequencies of the bilayer. This is done by setting IBRION=5 in the VASP calculations. It is worth to recall that within our supercell the matrix elements for $q$=0 wave vector can be mapped to the ($\frac{\pi}{a}(1,1)$) point of the single cubic-perovskite.}. 

As compared with the strongest electron-phonon coupling in the optimally doped bulk BBO, that is the breathing mode contribution (bBR)  \footnote{We have calculated the breathing mode contribution to the electron-phonon coupling of the doped bulk BBO ($\lambda_{bBR}$) within the same approximation as for $\lambda_{IST}$, obtaining the matrix elements from a frozen phonon approach along the full Brillouin zone. We've considered  a 2x2x2 FCC supercell, the doping effect was simulated by the Virtual Crystal Approximation, and the atomic displacements of the oxygens in the breathing mode was $0.044 \AA$, similarly as at the interface.}, we obtain $\lambda_{IST}\simeq\lambda_{bBR}/3$ for a wide range of gaussian smearings $\sigma$ values (0.002 Ry $\lesssim\sigma\lesssim$ 0.02 Ry) and well converged k-point grids. However, we remark that the absolute value that comes out of the double delta sum is sensitive to $\sigma$. The reported values for $\lambda_{bBR}$ and total $\lambda_{3D}$ for the optinally doped BBO are 0.3\cite{PhysRevB.44.5388,PhysRevB.57.14453} and 0.34\cite{PhysRevB.57.14453}, respectively. If we consider that total electron-phonon coupling at the interface of the smallest simulated slab is $\lambda_{2D}=\lambda_{3D}/3$, within PBE $\lambda_{2D}=0.11$.   

Resorting to the simplified expression
\begin{equation}
\lambda=N^{\uparrow}_0\ll g^2\gg/M\ll\omega^2\gg,
\end{equation}
where $\ll g^2\gg$ and $\ll\omega^2\gg$ are the properly averaged square of the coupling matrix elements and phonon frequencies as described in Ref. \citenum{McMilla1968} and, assuming that the frequencies do not significantly vary as compared to the bulk\footnote{Similar values for the breathing mode frequencies have been reported for bulk doped BBO\cite{PhysRevB.57.14453,Kotliar2013}.}, we can rescale the PBE value for $\lambda_{2D}$ of the thinner slab to the corresponding thicker one for the BPO4/BBO6 case withing HSE, following the method suggested in Ref. [\citenum{Kotliar2013}]: 

\begin{equation}
\lambda_{2D}^{HSE,thick}=\lambda_{2D}^{PBE,thin}\frac{N^{\uparrow}_{HSE,thick}\ll g^2 \gg_{HSE}}{N^{\uparrow}_{PBE,thin}\ll g^2\gg_{PBE}},
\end{equation}

where $N^{\uparrow}_{PBE,thin}$ is the calculated DOS at $E_F$ for the thinner BPO2/BBO4 slab within PBE that is equal to 1.25 states/eV and $N^{\uparrow}_{HSE,thick}$ is the corresponding one for the BPO4/BBO6 slab within HSE, equal to 2.5 states/eV. The rescaling of the averaged squared coupling matrix elements is done considering the calculated values 4.6 eV/$\AA$ (PBE)  and 5.9 eV/$\AA$(HSE), correspondingly. We recall that this coupling shows no sensitivity to the width of the BPO as mentioned before. By doing this rescaling we obtain that the $\lambda_{2D}^{HSE,thick}=0.36$. 

Considering the modified McMillan equation for the superconducting  critical temperature,
\begin{equation}
T_c=\frac{\omega_{log}}{1.2}exp\left( -\frac{1.04(1+\lambda)}{\lambda-\mu^*(1+0.62\lambda)}\right),   
\label{mcmillan}
\end{equation}
and taking $\omega_{log}$~=~450~K and $\mu^*$~=~0.1, the same parameters reported in Ref.~\onlinecite{Kotliar2013} for BBO bulk, our obtained value of $\lambda_{2D}$ gives rise to $T_c\sim$ 1K, that is of the same order of magnitude of the experimentally observed $T_c$ for the BPO/BBO bilayer.  Thicker BPO bilayers are expected to increase even more the DOS at $E_F$ and, concomitantly, enlarge $T_c$.

\begin{figure}[h]
\includegraphics[width=1.\columnwidth]{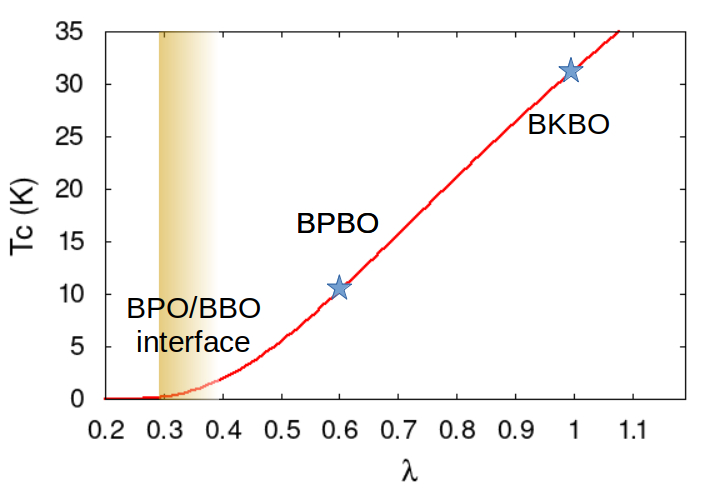}
\caption{Schematic evolution of $T_c$ for the BBO family of superconductors according to the McMillan modified equation. }.   
\label{fig:Tc}
\end{figure}

The main message of this work, then, is that there is a substantial coupling between the electrons and the interfacial stretching vibrations and a 
large $N^{\uparrow}_0$ that can explain the experimentally observed 2D superconductivity bellow 3.5 K. In Fig. \ref{fig:Tc}, we show a schematic 
representation of the evolution of $T_c$ along the BBO family of superconductors that can be qualitatively explained by Eq. \ref{mcmillan}. In 
Ref.[\citenum{Kotliar2013}], a $\lambda_{3D}\sim$ 1 was calculated within HSE and confirmed experimentally through ARPES\cite{PhysRevLett.121.117002}.
Interestingly, the decrease of $T_c$ from BKBO to BPBO can be obtained by just rescaling the DOS at $E_F$ ( going from 0.44 to 0.26 
states/eV \cite{PhysRevLett.60.2681}) while keeping $\ll g^2\gg$ fixed because the electronic structure remains similar. The system seems to bear 
lower level of doping when Pb substitutes Bi. Contrary to what happens at the BPO/BBO inerface, in the bulk cases, there is a small 
DOS at $E_F$ and the high $T_c$ has its origin in the large $\lambda$.  

In a previous work, it has been predicted by means of DFT calculations  that the (001) BBO surface should become metallic if a Bi-termination is achieved~\cite{Vildosola2013}. The mechanism behind this effect should be intrinsic and has to do with the breaking of charge ordering at the two outermost layers due to the incomplete oxygen environment of the surface ions. Moreover, in view of the observed electronic and structural reconstruction of BiO$_2$ surface, it's been claimed that the predicted 2D electron gas deserved further investigation regarding its superconducting properties. We believe that the physics at the BPO/BBO interface is related to the one described for the Bi-terminated (001) BBO surface. In both cases, there is an electronic and structural reconstruction due to a sudden change in the oxygen octahedra environment of the charge ordered Bi ions. 

In particular, at the BPO/BBO interface there is a partial suppression of the BBO charge order due to the breaking of the translation symmetry along the z direction by the presence of the BPO layer. There are two types of electronic states at the interface, the sp-Pb electron doping bands and the s-Bi ones. The former are more spatially extended while the last ones are 2D. In this work, we  show that the 2D interfacial states present a substantial electron-phonon coupling with the stretching modes and are responsible for the mechanism behind the experimentally observed 2D superconductivity at the BPO/BBO bilayer. These results should encourage future works to search different ways to break the charge ordered phases at interfaces with semiconductors like BBO, for instance, to involve also the $Bi^{+5}$ empty bands. We think that electrostatic doping is an interesting possibility to be explored.

The authors receive financial support from PIP2015-2017 112-201501-00364-CO of CONICET and from PICT 2015 0869, of the ANPCyT, Argentina.

\bibliography{biblio_gases2D-oxidos}
\end{document}


\title{Supplemenatry Information of "2D superconductidy driven by interfacial electron-phonon coupling at the BaPbO$_3$/BaBiO$_3$ bilayer
"}

\author{Solange Di Napoli, Christian Helman, Ana María Llois, and Verónica L. Vildosola}

\maketitle

\section{Computational details}

Calculations are performed using density functional theory (DFT) with the projector augmented wave method (PAW) as implemented in the Vienna Ab initio Simulation Package (VASP)~\cite{VASP,PAW-VASP}. In order to properly describe the electronic structure of charge ordered and semiconducting BaBiO$_3$, it is necessary to consider long range exchange interactions that are poorly described by the standard local and semilocal functionals as the Local Density Approximation or the General Gradient Approximation. As will be shown in the next section of this Supplementary material, we arrive to the conclusion that semilocal functionals as the Perdew-Burke-Ernzerhof (PBE)~\cite{PBE96} describe the ionic relaxions reasonably well. Then, for the simulated slabs, we use the hybrid functional including a fraction of non-local Hartre-Fock exchange interactions through the  Heyd-Scuseria-Ernzerhof (HSE06) functionals~\cite{2003-HSE1,2006-HSE2} only at the electronic level. The PBE and HSE calculations are performed using a 500~eV energy cutoff in the plane waves basis.  To evaluate the integrals within the BZ a 8$\times$8$\times$1 and a 6$\times$6$\times$1 Monkhorst-Pack $k$-point grids are employed for PBE and HSE calculations, respectively. The internal structural relaxations are performed until the forces on each ion are less than 0.01~eV/\AA. 

We have simulated slabs with different sizes and compositions notated as BPOi/BBOj. The supercell is built along the (001) direction, $i$ is the number of formula units of the BPO part and $j$ the corresponding number of the BBO one.
We have calculated the following cases: BPO2/BBO4, BPO2/BBO6, BPO2/BBO8 and BPO4/BBO6. The full Brillouin zone calculation of the matrix elements for the BPO2/BBO4 slab has been performed with the Wien2k code \footnote{P.Blaha, K. Schwarz, G. K. H. Madsen, D. Kvasnicka, J. Luitz, R. Laskowski, 
F. Tran and L. D. Marks, WIEN2k, An Augmented Plane Wave + Local Orbitals Program
for Calculating Crystal Properties (Karlheinz Schwarz, Techn. Universit\"at Wien, Austria), 2018. ISBN 3-9501031-1-2}

\section{Assesing the breathing distortions and charge disproportionation with the PBE and HSE functionals}

In Table \ref{tab:structure}, we present the results for the breathing distortions and charge disproportionation upon ionic relaxation for bulk BaBiO$_3$. We show the different BiO distances for both the Bi$^{+3}$ and Bi$^{+5}$ obtained using the PBE and HSE functionals, and the comparison with the available experimental data~\cite{Cox1976969}. The charges have been calculated through Bader analyses~\cite{Bader}.

The calculation for bulk BaBiO$_3$ using the PBE functional was done with a 10$\times$10$\times$8 k-mesh and a 520~eV energy cutoff in the plane waves basis, while a 8$\times$8$\times$6 k-mesh and an energy cutoff of 450~eV was considered when using HSE.

\renewcommand{\thetable}{S1}

\begin{table}[h]
\begin{center}
\caption{ Bi$^{+3}$-O and Bi$^{+5}$-O distances. d$_{\parallel}$ (d$_{\perp}$) means parallel (perpendicular) to $ab$-plane. The charge disproportionation $\Delta q$ is calculated through the corresponding Bader charges. ($^*$) In this case, the HSE06 functional is considered only at the electronic level.}
\label{tab:structure}
\begin{tabular}{@{}|c|ccccc|}
\hline
			& Exp.~\cite{Cox1976969} & PBE & $\Delta_{Exp-PBE}$ &HSE  &$\Delta_{Exp-HSE}$ \\
\hline
d$_{\parallel}$ (Bi$^{3+}$-O)    &      & 2.28  &        & 2.30 &       \\
d$_{\perp}$ (Bi$^{3+}$-O)        &      & 2.29  &        & 2.31 &       \\
$\langle$ d(Bi$^{3+}$-O) $\rangle$ & 2.28 &  2.28 &  0.00  & 2.31 & -0.03 \\
d$_{\parallel}$ (Bi$^{5+}$-O)    &      &  2.14 &        & 2.11 &       \\
d$_{\perp}$ (Bi$^{5+}$-O)        &      &  2.15 &        & 2.11 &       \\
$\langle$ d(Bi$^{5+}$-O) $\rangle$ & 2.12 & 2.14  & -0.02  & 2.11 & -0.01 \\
$\Delta$(d$_{Bi-O}$)             & 0.17 & 0.14  & -0.03  & 0.20 &  0.03 \\
$\Delta$ q(Bi$^{3+}$-Bi$^{5+}$)    &      & 0.35 &         & 0.53 (0.42$^*$) &   \\
\hline				
\end{tabular}
\end{center}
\end{table}

From these results we conclude that the description of the breathing distortions and charge disproportionation obtained with PBE is reasonable, and therefore, for the slab calculations we use the more computational demanding HSE only at the electronic level. 

\section{Electronic reconstruction for the different simulated slabs}

\renewcommand{\thefigure}{S1}
\begin{figure}[h]
\begin{tabular}{cccc}
\includegraphics[scale=0.3]{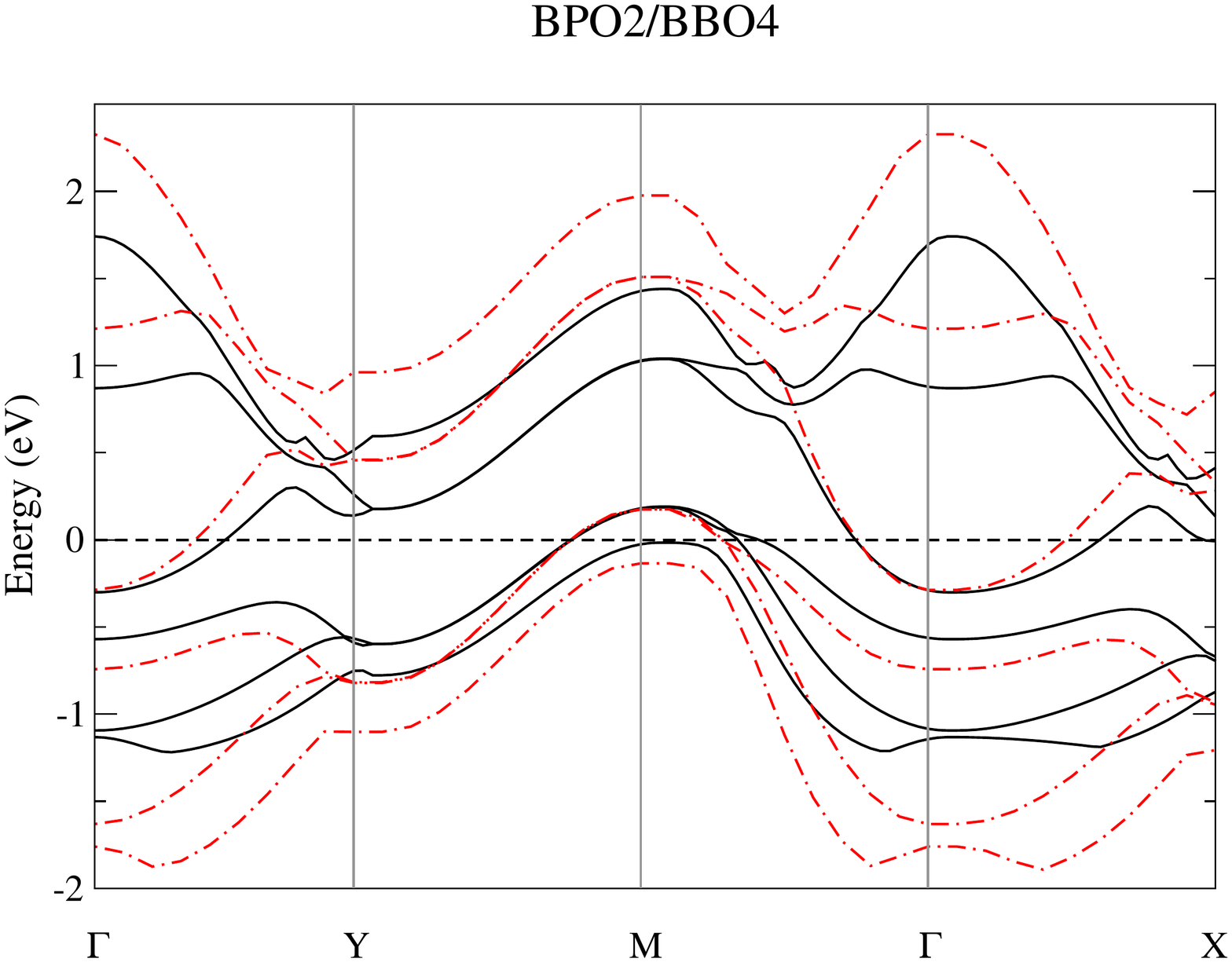}
& \includegraphics[scale=0.3]{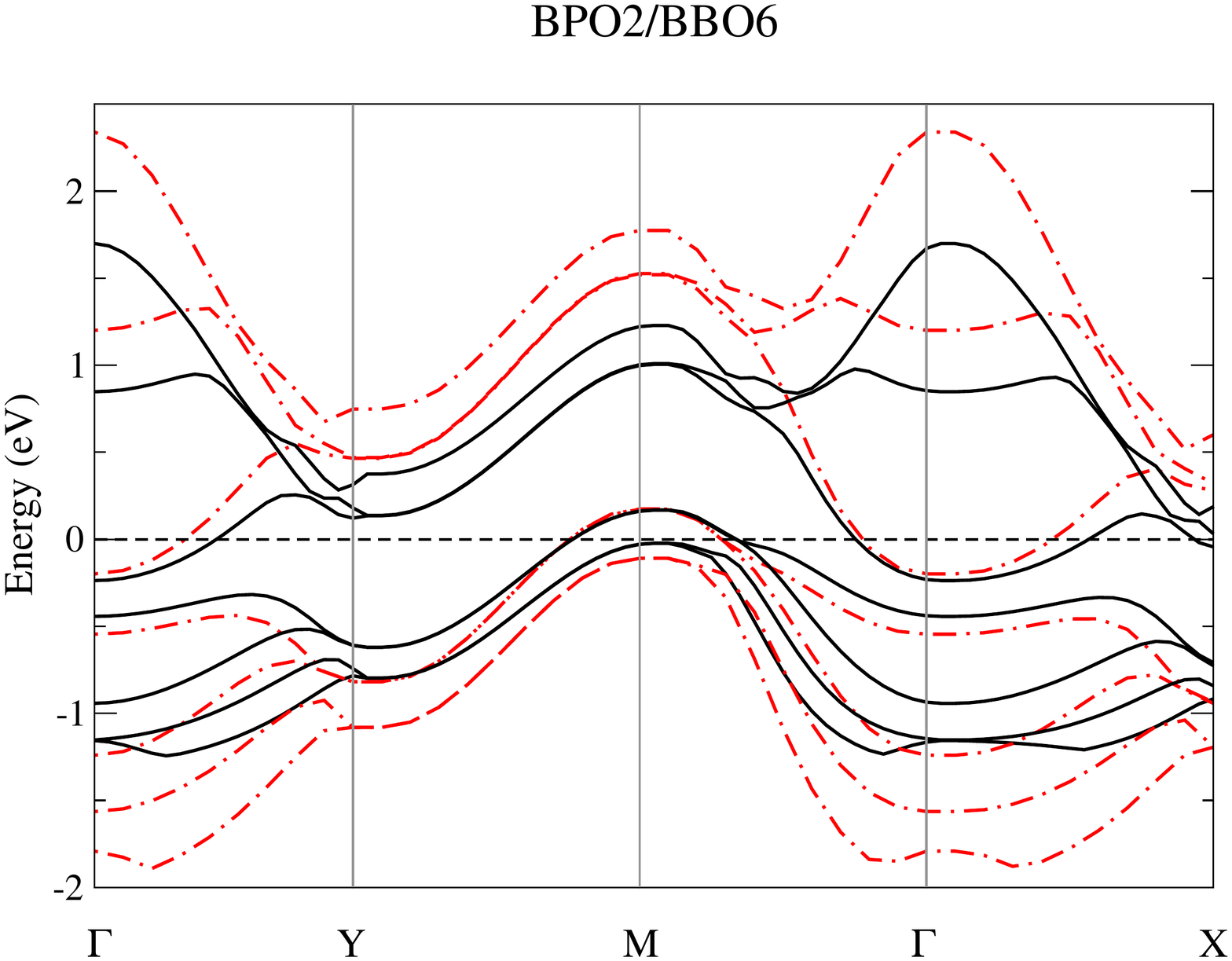}\\
\includegraphics[scale=0.3]{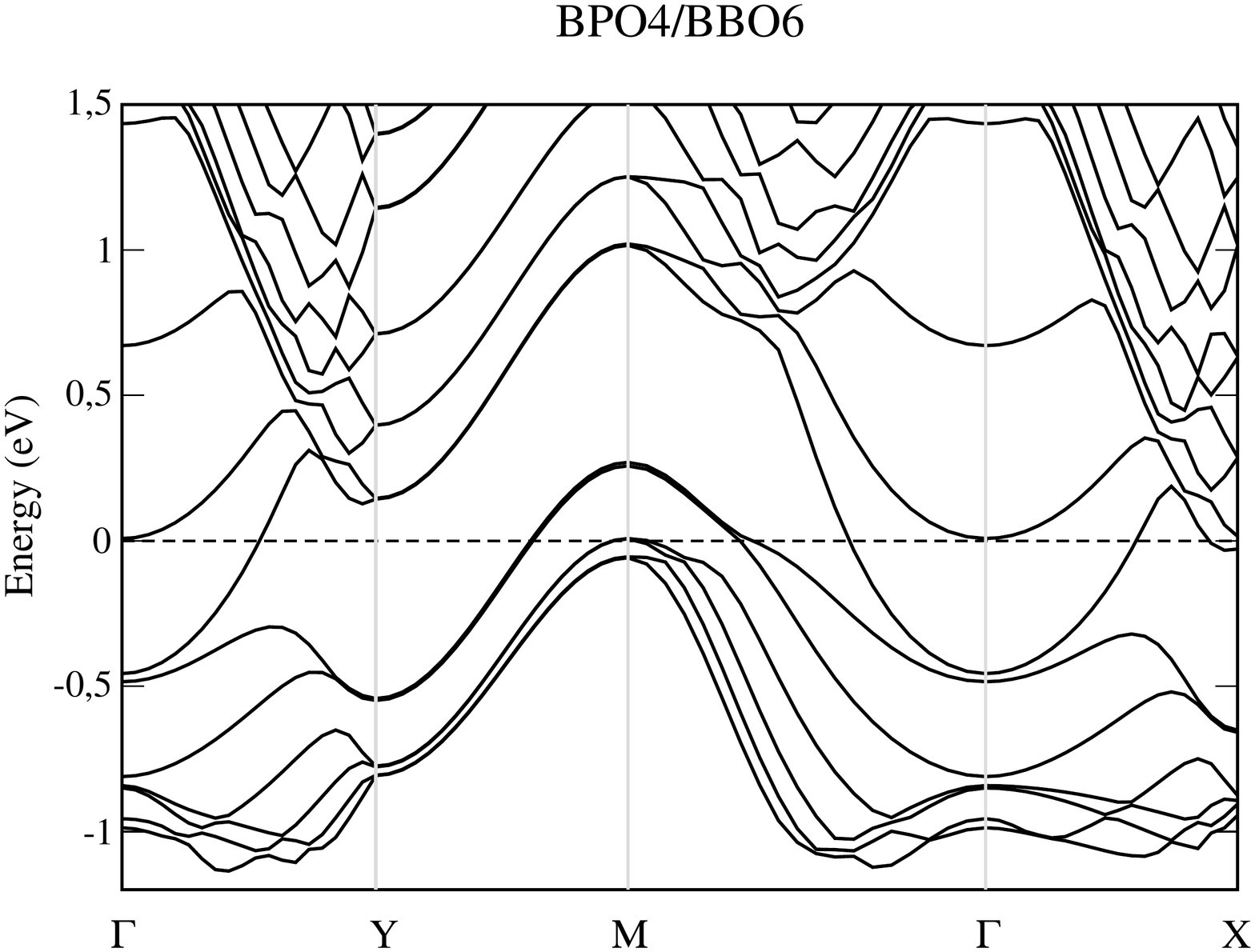}
&\includegraphics[scale=0.3]{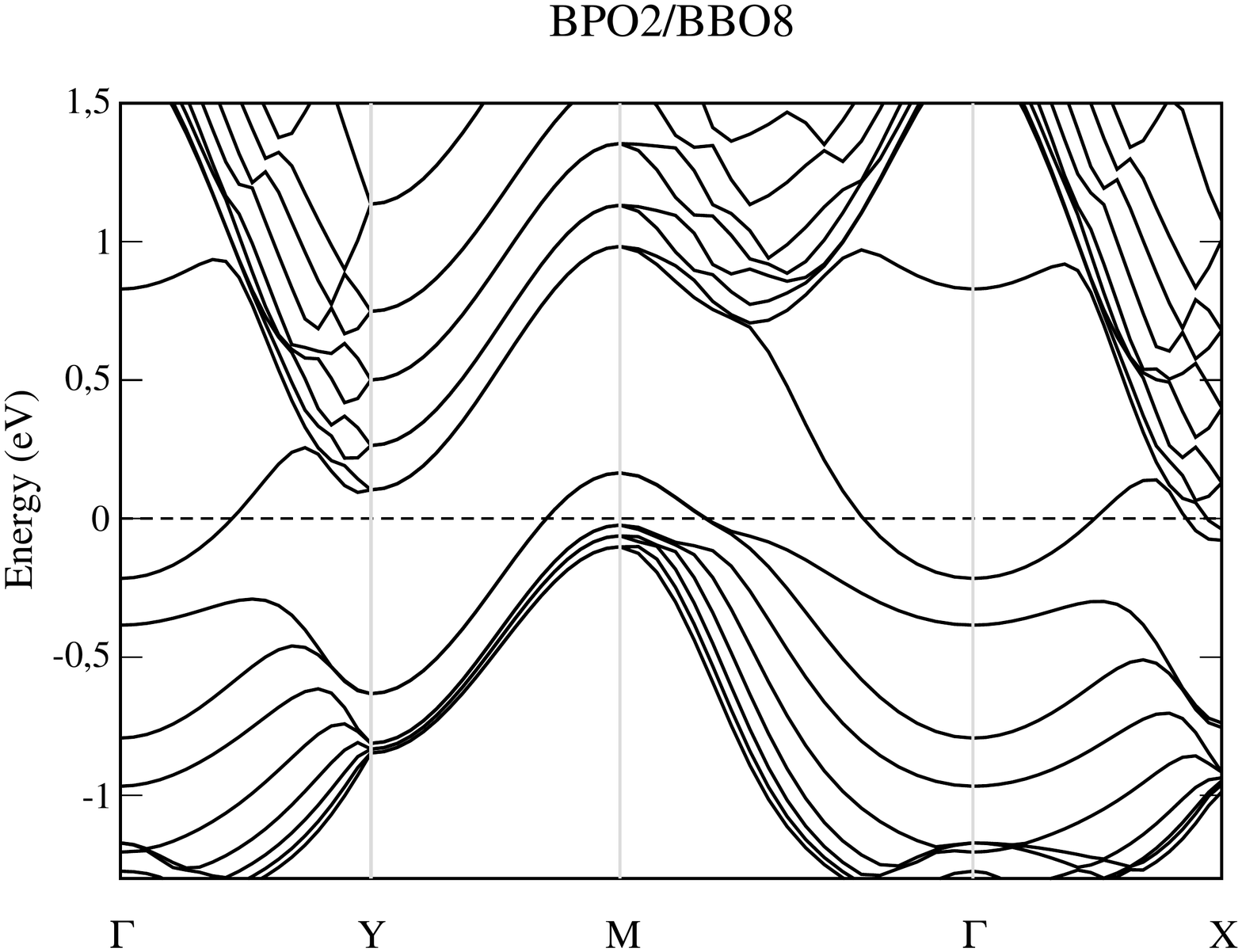}\\
\end{tabular}
\caption{Bandstructure for slabs with different size and composition. The PBE (HSE) bandstructure is depicted in black solid (red dash dot) line.}
\label{fig:bands}
\end{figure}

The smallest slab, BPO2/BBO4, contains the essential characteristics to describe the system, in the sense that the two innermost layers preserve the charge disproportionation, the breathing distortions and semiconducting behaviour expected for bulk BaBiO$_3$.

The hole pocket centered at the $M$ point, is not strongly influenced by the width of the BBO part. For instance, the height of the hole pocket does not significantly change for the BPO2/BBOm, with j = 4, 6 and 8. The corresponding energies are 0.19 eV, 0.17 eV and 0.16 eV respectively. On the other hand, when comparing the slabs BPOi/BBO6, with i=2 and 4, it can be observed that the size of this hole pocket increases with the number of BPO layers, since it is 0.26 eV for the BPO4/BBO6 case.

\section{ Assesing the role of the Spin-orbit coupling at the BPO/BBO bilayer}

We have performed calculations with the Wien2k code in order to asses the role of the Spin-Orbit (SO) coupling to the electronic structure of the BPOi/BBOj bilayer. In Figure S4, we show the calculated bandstructure for the BPO2/BBO4 slab obtained without (left) and with SO (right) coupling using the PBE functional. There are no appreciable changes for the states crossing the Fermi level that are the ones that are relevant in this work. The effect of the SO coupling is more important for those states that lie 2 eV above the Fermi level (remarked with a red ellipse). The topological properties of these states probably deserve further study but it is beyond the scope of the present work.

\renewcommand{\thefigure}{S2}
\begin{figure}[h]
\begin{tabular}{cc}
\includegraphics[scale=0.4]{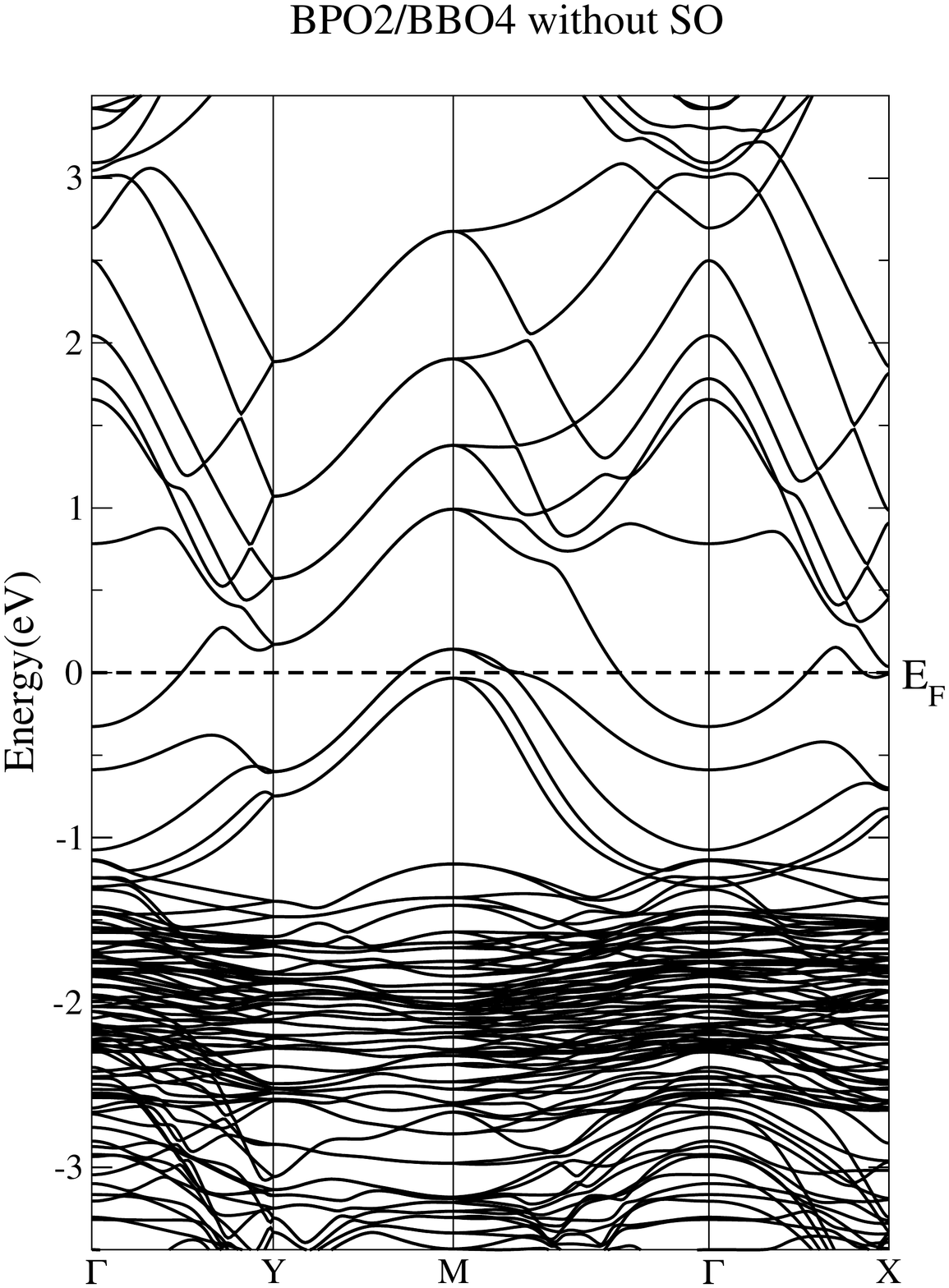}
\includegraphics[scale=0.4]{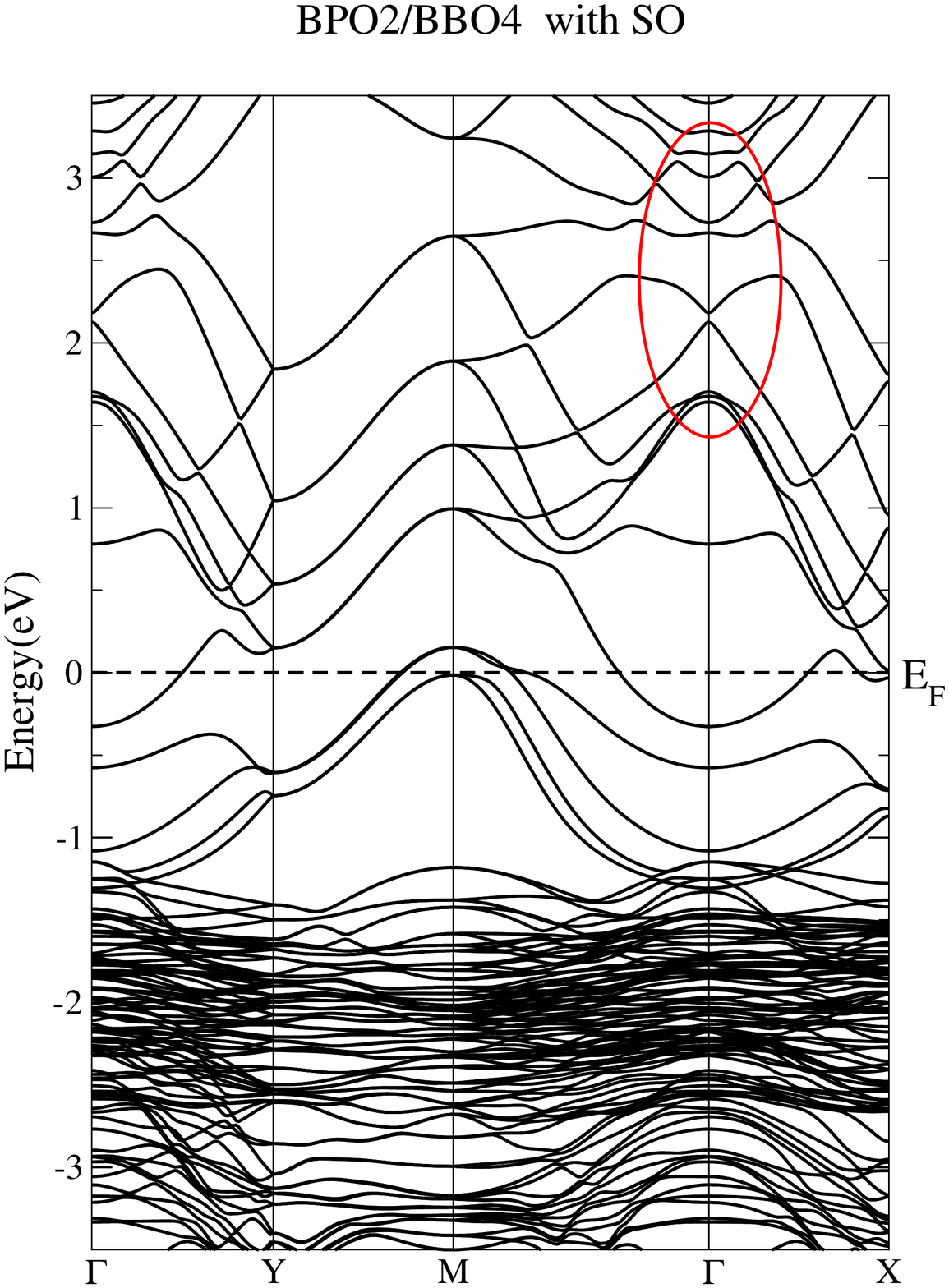}
\end{tabular}
\caption{Bandstructure for the BPO2/BBO6 slab obtained with the PBE functional without (left) and  with (right) SO coupling.}
\label{fig:bands}
\end{figure}

\section{Electron-phonon coupling}

The electron-phonon coupling $\lambda$ can be written as
\renewcommand{\theequation}{S1}
\begin{equation}
\lambda=\sum_{q\nu}\lambda_{q\nu}=\frac{2}{N^{\uparrow}_0}\sum_{nmk}\frac{|g^{\nu}_{nk,mk+q}|^2}{\omega_{q\nu}} \delta(\varepsilon_{nk}-E_F)\delta(\varepsilon_{mk+q}-E_F),
\label{eq:delta}\end{equation}

where $\omega_{q\nu}$ is the frequency of the phonon $\nu$ with wave vector $q$, $N^{\uparrow}_0$ is the DOS at $E_F$ per spin and $g^{\nu}_{nk,mk+q}$ are the electron-phonon matrix elements that are related to the self-consistent change of the Kohn-Sham potential under the phonon distortions. Following Refs. \citenum{PhysRevLett.55.837} and \citenum{PhysRevB.44.5388}, these matrix elements can be approximated through a frozen phonon approach by $g^{\nu}_{nk,mk+q}\sim\frac{1}{\sqrt(2M\omega_{q\nu})}<nk|(V_{q\nu}-V_0)/u_{q\nu}|mk+q>$, where $u_{q\nu}$ and $M$ are the displacement and mass of the ions involved in the phonon mode, while $V_{q\nu}$ and $V_0$ are the self-consistent potentials of the distorted and undistorted crystals, respectively. In this way, only intraband couplings ($n=m$) are considered and the corresponding matrix elements can be estimated at a given wave vector $q$ and mode $\nu$, directly from the shift of the bands calculated for a proper supercell, as long as  $q$ is commensurate with the lattice.

\subsection{Frozen phonon effect in bandstructure : PBE vs HSE }
 
 The shifts indicated with an arrow in each plot of Figure \ref{fig:HSE-bands}, $\Delta \varepsilon$, are 0.20 eV for PBE and 0.26 eV for HSE. The ionic displacements of the oxygens involved in the stretching mode is 0.044\,\AA. The shifts of the eigenvalues are larger close to the M point, as can also be observed at the full Brillouin zone plot of Figure \ref{fig:split-FBZ} in the next section. These shifts give rise to an estimation of the electron phonon matrix elements $\Delta \varepsilon/\Delta u$ of 4.6 eV/\AA~ for PBE and 5.9 eV/\AA~ for HSE. 
 
  \renewcommand{\thefigure}{S3}
 \begin{figure}[h]
 \includegraphics[scale=0.6]{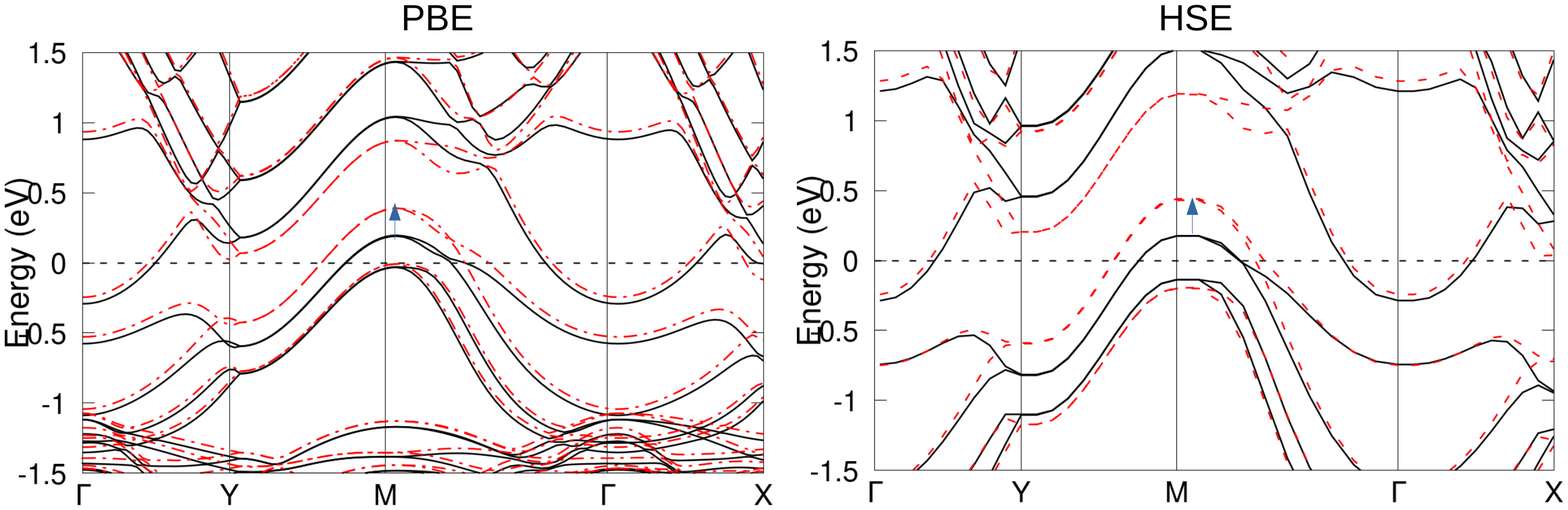}
 \vspace{-6cm}
 \caption{Bandstructures of the BPO2/BBO4 slab for the undistorted crystal. In dashed red line, the corresponding frozen phonon bandstructures under the interfacial stretching distortion.}
\label{fig:HSE-bands}
\end{figure}
 
 In Figure \ref{fig:PBE-FE}, we show that there are no appreciable changes in the bandstructure under the ferroelectric (FE) distortion for the Bi ions at the interface. The scheme of the distortion is depicted in the inset. The oxygen atoms's displacements considered are 0.044~\AA\, and the Bi's ones -0.0044 \AA; the center of mass is preserved.
 
 \renewcommand{\thefigure}{S4}
 \begin{figure}[h!]
\includegraphics[scale=0.4]{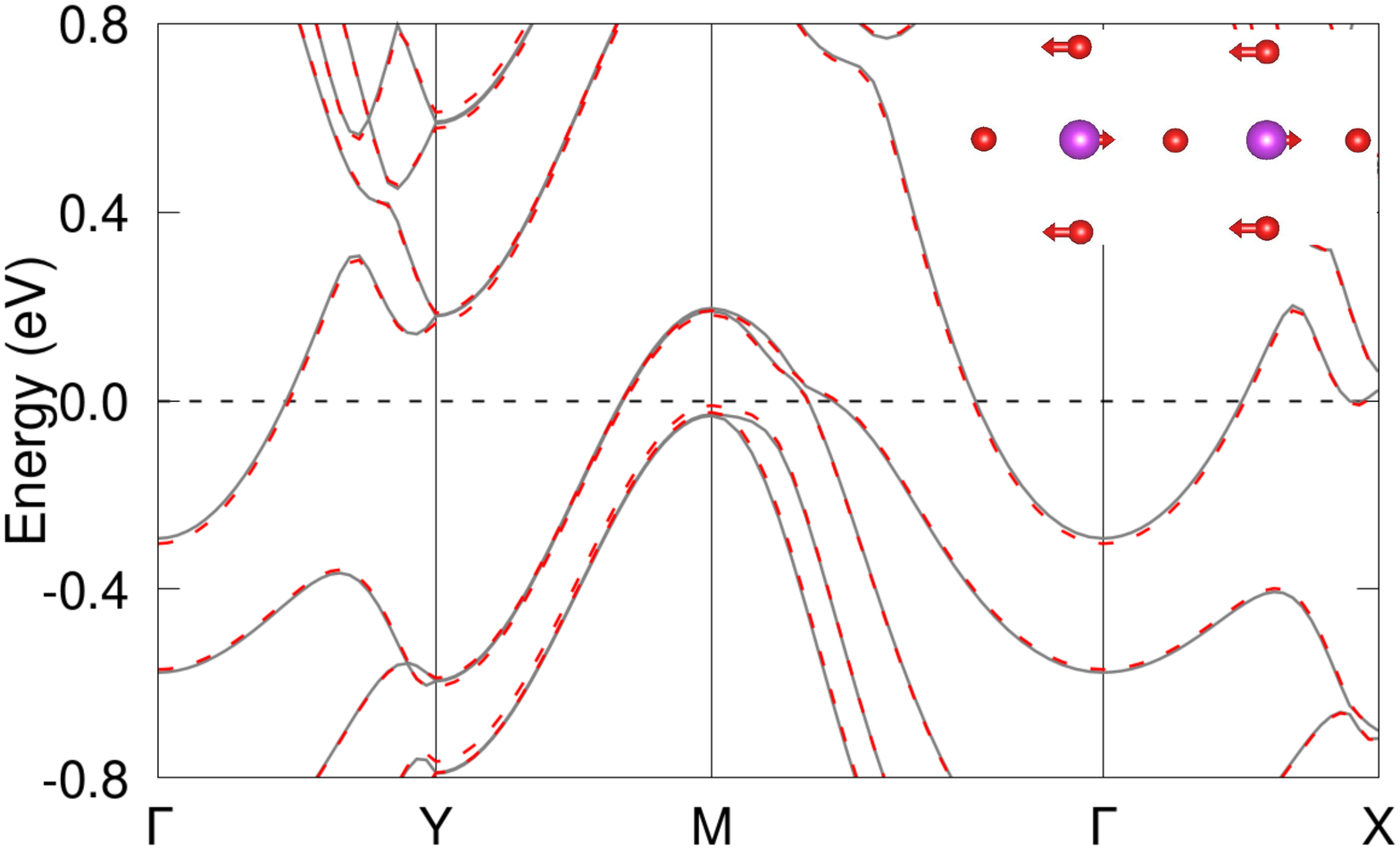} 
\vspace{-1cm}
\caption{In solid black line, the PBE bandstructure of the BPO2/BBO4 slab for the undistorted crystal. In dashed red line, the frozen phonon bandstructures under the interfacial ferroelectric distortion. In the inset, the  ferroelectric frozen phonon displacements are shown.}
\label{fig:PBE-FE}
\end{figure}

\subsection{Electron-phonon coupling along the full Brillouin zone }

\renewcommand{\thefigure}{S5}
\begin{figure}[h!]
\includegraphics[scale=0.4]{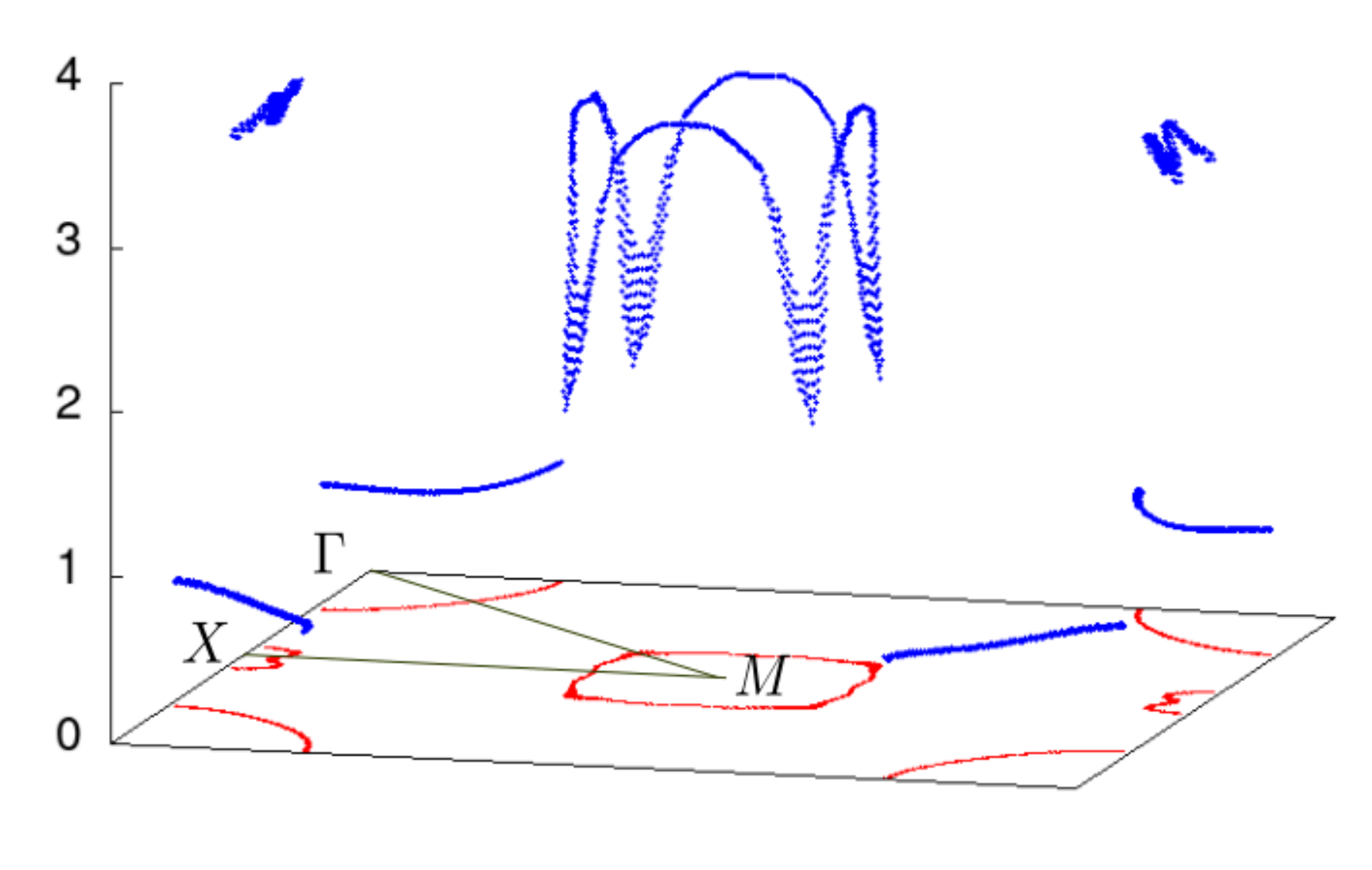}
\caption{Approximated electron-phonon matrix element (in eV/\AA ) for the interfacial stretching vibration mode along the full Brillouin zone of the BPO2/BBO4 heterostructure. At the base (z=0 plane) the Fermi surface is projected in red.}
\label{fig:split-FBZ}
\end{figure}

We calculate the approximated electron matrix elements of the interfacial ST mode along the full Brillouin zone for a dense k-point grid up to 100x100, using the PBE functional. In Fig. \ref{fig:split-FBZ} we show how these matrix elements depend on k. The Fermi surface is projected at the z=0 plane. There is one pocket around $\Gamma$ where there is a small contribution to the electron-phonon coupling. The strongest contribution comes, as expected, from the pocket centered at M where the electron-phonon coupling is quite anisotropic, being stronger along the M-X and M-Y directions than for the M-$\Gamma$ one. There is a tiny pocket at X that is an artifact of PBE because within HSE it is above E$_F$, as it can be seen in Fig. \ref{fig:HSE-bands}.      

\bibliography{References-BPO-BBO}